\documentclass[preprint2]{aastex} 

\usepackage{natbib}
\slugcomment{Accepted by ApJ, June 2014}

\shorttitle{3C\,220.3}
\shortauthors{Haas et al.}

\newcommand{\Lsol}{L$_{\odot}$}
\newcommand{\Msol}{M$_{\odot}$}
\newcommand{\0}{\phantom{0}}  
\newcommand{\HST}{{\it HST\/}}
\newcommand{\Hubble}{{\it Hubble Space Telescope\/}}
\newcommand{\Her}{{\it Herschel\/}}
\newcommand{\HSO}{{\it Herschel Space Observatory\/}}
\newcommand{\SST}{{\it Spitzer Space Telescope\/}}
\newcommand{\Sp}{{\it Spitzer\/}}

\begin{document}


\title{3C\,220.3: a radio galaxy lensing a submillimeter galaxy}



\author{
  Martin Haas\altaffilmark{1}, 
  Christian Leipski\altaffilmark{2}, 
  Peter Barthel\altaffilmark{3},  
  Belinda J. Wilkes\altaffilmark{4}, 
  Simona Vegetti\altaffilmark{5}, 
  R. Shane Bussmann\altaffilmark{4}, 
  S. P. Willner\altaffilmark{4},
  Christian Westhues\altaffilmark{1}, 
  Matthew L. N. Ashby\altaffilmark{4},
  Rolf Chini\altaffilmark{1,6}, 
  David L. Clements\altaffilmark{7}, 
  Christopher D. Fassnacht\altaffilmark{8}, 
  Assaf Horesh\altaffilmark{9}, 
  Ulrich Klaas\altaffilmark{2}, 
  L\'eon V. E. Koopmans\altaffilmark{3}, 
  Joanna Kuraszkiewicz\altaffilmark{4}, 
  David J. Lagattuta\altaffilmark{10,11}, 
  Klaus Meisenheimer\altaffilmark{2}, 
  Daniel Stern\altaffilmark{12}, 
  Dominika Wylezalek\altaffilmark{12,13}
}

\altaffiltext{1}{ Astronomisches Institut, Ruhr Universit\"at, Bochum, Germany, 
  email: haas@astro.rub.de}
\altaffiltext{2}{ Max-Planck-Institut f\"ur Astronomie, Heidelberg, Germany}
\altaffiltext{3}{ Kapteyn Astronomical Institute, University of Groningen, The Netherlands} 
\altaffiltext{4}{ Harvard-Smithsonian Center for Astrophysics, Cambridge, Massachusetts, USA}
\altaffiltext{5}{ Max-Planck-Institut f\"ur Astrophysik, Garching, Germany} 
\altaffiltext{6}{ Universidad Catolica del Norte, Antofagasta, Chile}
\altaffiltext{7}{ Imperial College, London, UK}
\altaffiltext{8}{ University of California, Davis, USA}
\altaffiltext{9}{ Division of Physics, Mathematics, and Astronomy, California Institute of Technology, Pasadena, California, USA} 
\altaffiltext{10}{ Centre for Astrophysics \& Supercomputing, Swinburne University of Technology, Hawthorn, Australia} 
\altaffiltext{11}{ ARC Centre of Excellence for All-sky Astrophysics (CAASTRO)} 
\altaffiltext{12}{ Jet Propulsion Laboratory, California Institute of Technology, Pasadena, California, USA}
\altaffiltext{13}{ European Southern Observatory, Garching, Germany}


\begin{abstract}
  \HSO\ photometry and extensive multiwavelength followup have
  revealed that the powerful radio 
  galaxy 3C\,220.3 at $z=0.685$  acts as a gravitational lens for a background 
  submillimeter galaxy (SMG) at $z=2.221$. 
  At an observed wavelength of 1~mm, the SMG
  is lensed into three distinct images.  In the observed 
  near infrared, these images are connected by an arc 
  of $\sim$1\farcs8 radius forming an Einstein half-ring 
  centered near the radio galaxy. 
  In visible light, only the arc is apparent.
  3C\,220.3 is the only known instance of strong galaxy-scale 
  lensing by a 
  powerful radio galaxy not located in a galaxy cluster and therefore it
  offers the potential to probe the dark matter content 
  of the radio galaxy host.
  Lens modeling rejects a single lens, but two lenses centered on 
  the radio galaxy host A and a companion B, separated by 1\farcs5,  
  provide a fit consistent with all data and reveal
  faint candidates for the predicted fourth and fifth images. 
  The model does not require an extended common dark matter halo,
  consistent with the absence of extended bright X-ray emission 
  on our {\it Chandra} image. The 
  projected 
  dark matter fractions 
  within the Einstein radii of A (1\farcs02) and B (0\farcs61) 
  are about 0.4$\pm$0.3 and 0.55$\pm$0.3.
  The mass to $i$--band light ratios of A and B, 
  $M/L_{i} \sim 8 \pm 4~M_{\odot}/L_{\odot}$, 
  appear comparable to those of radio-quiet lensing
  galaxies at the same redshift in the CASTLES, LSD, and SL2S samples.
  The lensed SMG is extremely 
  bright with observed $f({250\,\micron}) = 440$~mJy 
  owing to a magnification factor 
  $\mu \sim 10$.  
  The SMG spectrum shows luminous, narrow 
  \ion{C}{4} $\lambda$1549\,\AA\ emission, revealing that the 
  SMG houses a hidden 
  quasar in addition to a violent starburst. 
  Multicolor image reconstruction of the SMG indicates a bipolar 
  morphology of the emitted ultraviolet (UV) light 
  suggestive of cones through which UV light 
  escapes a dust-enshrouded nucleus. 
\end{abstract}




\section{Introduction}

Powerful radio galaxies (PRGs, with 
$L_{\rm 178\,MHz} > 10^{27}$\,W\,Hz$^{-1}$\,sr$^{-1}$
and Fanaroff-Riley morphology type FR~II) mark places of 
exceptional energy transformation. Giant radio lobes far 
outside the massive host galaxies are powered by 
relativistic jets launched from accreting supermassive 
black holes in the galaxy centers. 
PRGs are among the most massive galaxies at their epoch, 
as traced by the $K$--$z$ brightness--redshift diagram 
(\citealt{Lilly1984}). 
In the cold dark matter paradigm of cosmic structure formation 
(\citealt{White1978}), 
PRGs might be expected to reside in the most massive dark 
matter halos and in rich galaxy clusters.  
For a few PRGs in rich, X-ray-luminous clusters  
massive amounts of dark matter have been revealed by giant 
luminous arcs, e.g., 3C\,220.1 at $z = 0.61$ (\citealt{Dickinson1993}) 
or by the weak shear of background galaxies, e.g.,  
3C\,324 at $z = 1.2$ (\citealt{Smail1995}), 
both caused by cluster-scale gravitational lensing. 
However, for PRGs not in a cluster, no galaxy-scale 
gravitational lensing measurements of their dark 
matter content have been achieved to date. 

This paper reports the first case of a lensing PRG
not located in a galaxy cluster. 
As a bonus, the lensed source is a submillimeter galaxy 
(SMG)\null. SMGs are dust-enshrouded 
star-forming galaxies at high redshift ($z > 1$) and are 
thought to be the progenitors of today's massive 
elliptical galaxies. Lensed examples of SMGs are not rare 
(\citealt{Negrello2010}), 
but a full picture of the population has yet to emerge.

Our discovery of the lensing nature of 3C\,220.3 
is based on far infrared and 
submillimeter photometry with 
the \HSO.
The source appeared abnormally bright,
suggesting contamination of a background 
source in the \Her\ beam (5$\arcsec$ -- 35$\arcsec$).
Reprocessed archival \Hubble\  ({\HST}) 
visible (702\,nm) images revealed an arc surrounding 
two objects, 
indicating that gravitational lensing plays a role.
To identify the radio source among the two lensing 
candidate galaxies, we obtained
a deep 3.3\,cm radio map with the Karl G. Jansky 
Very Large Array (JVLA, 0\farcs4 beam width).
We performed 
high resolution imaging at 1.0\,mm with the 
Submillimeter Array (SMA), near-infrared 
(NIR, $K^{\prime}$-band, 2.124\,$\mu$m) adaptive optics (AO) 
imaging with the 10\,m Keck\,II telescope, 
3.6 and 4.5\,$\mu$m imaging with the 
{\it Spitzer Space Telescope\/},
360--880\,nm spectroscopy 
with the 5\,m Palomar telescope, 320--1000\,nm 
spectroscopy with the 10\,m Keck\,I telescope, and obtained 
a  10\,ks {\it Chandra\/} X-ray image. 

This paper reports results of the imaging and spectroscopy campaigns
(Section~\ref{section_observations_and_data}) 
and of lens modeling (Section~\ref{section_lens_modeling}) to fit the
resulting data.  Section~\ref{section_smg_reconstruction} 
treats the wavelength dependent morphology of the SMG.
Section~\ref{section_mass_to_light_ratio} 
compares the derived mass to light ($M/L$)
ratio for the PRG to ratios for galaxies at similar redshift, and
Section~\ref{section_conclusions} summarizes the results.
We adopt a standard $\Lambda$CDM cosmology 
($H_{\circ} = 72$\,km\,s$^{-1}$\,Mpc$^{-1}$, $\Omega_{\Lambda}=0.73$, 
and $\Omega_{m}=0.27$).

\section{Observations and data}
\label{section_observations_and_data}

\subsection{\HSO}

Photometric observations of 3C\,220.3 were carried 
out with the \HSO\
(\citealt{Pilbratt2010}) using the 
70, 100, and 160\,$\mu$m channels of the 
instrument PACS (\citealt{Poglitsch2010}) 
on UT 2011 May 28 in scan-map mode with 
on-source time of $\sim$80\,s at 70 and 100 $\mu$m
and $\sim$160\,s at 160\,$\mu$m.  
SPIRE (\citealt{Griffin2010}) 
observations at 250, 350, and 500\,$\mu$m 
were made on UT 2012 November 5 in small-map 
mode with $\sim$111\,s on-source time. 
Data reduction was performed within the 
Herschel Interactive Processing Environment 
(HIPE, \citealt{Ott2010})  
following standard procedures including 
source masking and high-pass filtering. 
Flux densities were measured for PACS with 
aperture photometry including corrections 
for aperture size. SPIRE photometry utilised
the timeline source extractor implemented in 
HIPE\null. Because the source is bright 
in all filters, the photometric uncertainties 
are small, typically a few percent. 

3C\,220.3 was observed with the MIPS (\citealt{Rieke2004}) 
instrument on the {\it Spitzer Space Telescope} 
(\citealt{Werner2004}) 
with reported 70 and 160\,$\mu$m flux densities 
(\citealt{Cleary2007}) significantly lower than 
those from \Her. To understand this discrepancy, 
we retrieved newly calibrated MIPS maps from the 
\Sp\/ archive and performed photometry with 
aperture corrections from the MIPS Instrument 
Handbook.\footnote{http://irsa.ipac.caltech.edu/data/SPITZER/docs/mips/
  mipsinstrumenthandbook/50}
The new 70 and 160\,$\mu$m \Sp\/ photometry yielded higher 
flux densities which now are consistent within 15\% with 
the \Her\ results. 
The \Sp\/ MIPS 24\,$\mu$m photometry 
agrees to within 4\% with 22\,$\mu$m photometry from 
the Wide-field Infrared Survey Explorer ({\it WISE\/}, 
\citealt{Wright2010}).

For comparison with 3C\,220.3, a sample of 12 PRGs was
taken from the 3CR catalog.  The sample (Westhues et al., in
preparation) consists of 
galaxies with $0.6 < z < 0.8$ and radio and 
mid-infrared properties similar to those of 
3C\,220.3.
The MIPS
24\,$\mu$m photometry of galaxies in the comparison 
sample is consistent with the {\it WISE\/} data.

\subsection{Karl G. Jansky Very Large Array}

\begin{figure*}
  \includegraphics[width=16cm,clip=true]{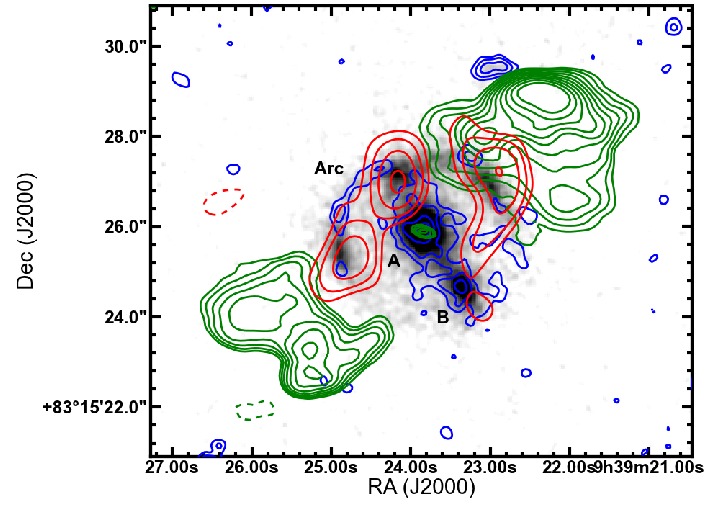}
  \includegraphics[width=16cm,clip=true]{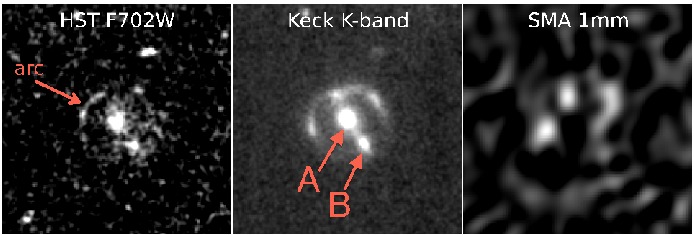}
  \caption{
    {\it Top:} negative grey-scale Keck AO ($K^{\prime}$-band, 2.124\,$\mu$m) 
    image of 3C\,220.3 with linear contours at other 
    wavelengths: blue = \HST\ at 702\,nm; 
    red = SMA at 1\,mm; green = JVLA at 3.3\,cm. 
    The position of the radio core is at 
    RA=09:39:23.837, Dec=+83:15:25.90 (J2000).
    {\it Bottom:} High resolution \HST, Keck AO, and SMA images of 
    3C\,220.3 with key components labeled. 
    Image sizes are 12$\arcsec$ square with north up and east to the left. 
  }
  \label{fig_high_res_images}
\end{figure*}

The deepest radio image of 3C\,220.3 to date, 
at $\lambda$ = 6\,cm by \citet{Mullin2006}, 
shows the double-lobed morphology 
but not the unresolved core.
With the aim of detecting the core, we observed 
3C\,220.3 with the 
Karl G. Jansky Very Large Array (JVLA) 
in its long-baseline A configuration 
on UT 2012 November 30 employing 25 antennae. 
The bandwidth was 2048\,MHz centered at 9\,GHz (X-band). 
Prime flux calibrators were 3C\,48 and 3C\,147, and the nearby
radio source J1010+8250 served as phase calibrator. The total
on-source integration time was 37~min. 

The resulting image, which 
was made using CASA and employed natural weighting 
of the visibility data, 
has an angular resolution of 
$\sim$0\farcs4. It shows the $\sim$9$\arcsec$ 
double-lobed morphology reported previously 
(\citealt{Mullin2006}) and an unresolved core with 
a flux density of 0.8\,mJy (Fig.~\ref{fig_high_res_images}). 

\subsection{\Hubble}

The \HST\ archive 
contains two images of 3C\,220.3 with WFPC2
in the broad band filter F702W taken on 
1994 March 11 and 1995 May 19, 
each with an exposure time of 300\,s. 
The reprocessed preview images now in the archive\footnote{
  http://archive.stsci.edu/} clearly show an arc, 
which was not visible on the early map published by 
\citet{McCarthy1997}.\footnote{\citet{McCarthy1997}, 
  reporting on early {\it HST} imaging of 3C radio galaxies, 
  mentioned the possible association of 3C\,220.3 with a 
  gravitational arc, citing ''M. Dickinson, private communication''. 
  While that claim now fortuitously turns out to have been correct, 
  the comment was at the time meant to apply to 3C\,220.1 
  (M. Dickinson, private communication). Dickinson has 
  further confirmed that the ground-based imaging of 
  3C\,220.3 available to him at that time was quite poor, 
  and 3C\,220.3 was one of the few targets omitted from his 
  comprehensive 3C optical imaging survey (\citealt{Dickinson1994}).  
  The 3C\,220.1
  lensing arc is 9\arcsec~in radius, 
  but the lens is an entire cluster
  (\citealt{Ota2000}, \citealt{Comerford2006}, \citealt{Belsole2007}). } 
To improve the contrast further, we have retrieved, 
aligned, and coadded the two F702W images
(Fig.~\ref{fig_high_res_images}), adjusting 
the positions of A and B to match those from the Keck AO 
(see Sect.~\ref{section_keck}). 

Photometric calibration was derived by measuring 
the total 702\,nm flux density in a circular 
aperture of 5\farcs6 diameter, containing sources 
A and B and the entire arc, and scaling it to the 
total $r$ and $i$ band photometry of 
3C\,220.3 listed 
by the Sloan Digital Sky Survey (SDSS, Data 
Release~9)\footnote{http://www.sdss3.org/dr9/}.
The total magnitudes (in the AB system) 
are $r = 20.96 \pm0.13$ at 623\,nm 
and $ i = 20.09 \pm 0.10$ at 764\,nm, 
yielding the total 702\,nm flux density listed in 
Table~\ref{table_photometry}. 
The 702~nm flux densities of A and B  separately are listed 
in Table~\ref{table_a_b}. 

\subsection{Near-infrared imaging with Keck adaptive optics}
\label{section_keck}

The 3C\,220.3 system was observed with the Keck~II 
telescope on UT 2012 December 24 utilizing NIRC2 
behind the adaptive optics bench. 
The laser guide star system was used in conjunction 
with an $R = 15.5$\,mag tip-tilt star located 
23$\arcsec$ from 3C\,220.3. 
The wide camera was used, which provides a pixel 
scale of 0\farcs0397 and a field of view of 40$\arcsec$ 
on a side. We obtained nine dithered exposures in the 
$K^{\prime}$~band (2.124\,$\mu$m), 
each exposure consisting of four co-added frames of 30\,s each. 

The data were reduced using a Python pipeline 
that flat-fields the data, subtracts sky emission, 
aligns the images, and drizzles to create the final image. 
The last step corrects for distortions that 
create a non-uniform pixel scale. 
The final image (Fig.~\ref{fig_high_res_images}) 
used eight of the nine input images, 
giving a total exposure time of 960\,s. 
The image center position was derived from three 
2MASS point sources in the frame. 
With this astrometry, A is within 0\farcs4
of the radio core position, and the final image 
coordinates were adjusted by this amount to match A 
to the radio core position 
(Fig.~\ref{fig_high_res_images}).

Photometric calibration was derived 
using an object (probably a compact galaxy)
located on the Keck AO mosaic
at RA = 09:39:34.18, Dec = +83:14:51.4 (J2000), and  
$K_{s} = 15.04 \pm 0.124$\,mag listed by 2MASS.

\subsection{\Sp/IRAC imaging}
\label{section_spitzer}

We observed 3C\,220.3 on UT 2013 June 14 
using the IRAC instrument 
(\citealt{Fazio2004})
of the \SST\
(\citealt{Werner2004}).
Nine dithered exposures of 30\,s each were secured 
at 3.6\,$\mu$m and at 4.5\,$\mu$m.
The dithered images were combined and resampled  to a pixel
size  of 0\farcs3 using IRACproc (\citealt{Schuster2006})
and MOPEX.

3C\,220.3 is clearly resolved at both 3.6 and 4.5\,$\mu$m. 
Approximate deconvolution gives full widths at half maximum surface
brightness  of 3\farcs0
at 3.6\,$\mu$m and 3\farcs4 at 4.5\,$\mu$m.  
These are comparable to the diameter of the Einstein ring (3\farcs6). 
However, the limited angular resolution makes 
it not straightforward to distinguish  
the components seen on the KECK AO image, i.e., 
sources A and B and the knots and arc.
The IRAC photometry of 3C\,220.3 listed in 
Table~\ref{table_photometry} 
is consistent with that from {\it WISE}.

\subsection{Submillimeter Array}

SMA imaging of 3C\,220.3 was obtained with 
Director's Discretionary Time in the 2012B 
semester as part of program 2012B-S084. 
Observations were conducted in extended array 
configuration (maximum baseline length of 226\,m) 
on UT 2013 January 11 in good weather conditions 
($\tau_{\rm 225~GHz} \sim 0.08$) for $t_{\rm int} = 9.6$~hours on-source
integration time.
Compact array observations (maximum baseline 
length of 76\,m) were obtained on 
UT 2013 February 10, again in good weather 
conditions ($\tau_{\rm 225~GHz} \sim 0.08$) 
for $t_{\rm int}  = 6.7$~hours. 
Phase stability for both observing nights 
was good (phase errors between 10\arcdeg\ and 20\arcdeg\ rms). 
The SMA receivers provided 8\,GHz of instantaneous 
bandwidth (considering both sidebands) 
and were tuned such that the upper sideband was 
centered on 302.927\,GHz.

Calibration of the $uv$ visibilities was performed 
using the Interactive Data Language (IDL) MIR package. 
The blazar 3C\,84 was used as the primary bandpass 
calibrator and Titan for absolute 
flux calibration. The nearby quasars 0721+713 
($S_{\rm 303~GHz} = 5.5$\,Jy, 13\fdg6 from target) 
and 1048+717 ($S_{\rm 303~GHz} = 0.8$\,Jy 
12\fdg0 from target) were used for phase and 
gain calibration. 
We checked the reliability of the phase-gain 
solutions using the quasar 1044+809 
($S_{\rm 303~GHz} = 0.3$\,Jy, 3\fdg2 from target). 

The Multichannel Image Reconstruction, Image Analysis, 
and Display (MIRIAD) software package (\citealt{Sault1995}) 
was used to invert the $uv$ visibilities and deconvolve 
the dirty map. 
Natural weighting was chosen to obtain maximum 
sensitivity and resulted in an elliptical Gaussian 
beam with a full-width half-maximum (FWHM) of 
1\farcs40$\times$1\farcs17 and major axis 
position angle  146\arcdeg\ east of north.

\subsection{Visible Spectroscopy}

\subsubsection{Palomar}

We observed 3C\,220.3 for 1 hour using the 
Double Spectrograph on the Palomar 5\,m 
Hale Telescope on UT 2013 February 12. 
The observations were obtained with three  
1200\,s exposures at a position angle 
of 31\arcdeg\ and the 5500\,\AA\ dichroic, 
providing a wavelength coverage of 
3200--8800\,\AA. 
Conditions were photometric, though the seeing 
was poor ($>$2\arcsec). 
The observations used the 1\farcs5-wide 
longslit, the 600/4000 blue grating 
(resolving power $R \sim 1000$), and the 600/10000 
red grating ($R \sim 2400$). 
We processed the data using standard techniques 
and flux calibrated the spectra using standard 
stars from \citet{Massey1990} observed 
during the same night.

The Palomar observations failed to detect the 
arc but confirmed the redshift of 3C\,220.3 at 
$z = 0.685$ based on the detection of narrow 
\ion{C}{2}] $\lambda$2326, [\ion{O}{2}] $\lambda$3727, and 
  [\ion{O}{3}] $\lambda$5007 emission lines, typical 
  of powerful radio galaxies (\citealt{McCarthy1993}, 
  \citealt{Stern1999}). 
  The only previously published redshift of 3C\,220.3 
  to our knowledge was based on [\ion{O}{2}] $\lambda$3727 
  and [\ion{Ne}{5}] $\lambda$3346 emission with no figure 
  available (\citealt{Spinrad1985}).\footnote{Jasper Wall 
    kindly showed us a plot of another spectrum of 3C\,220.3 taken 
    with the 4\,m William Herschel Telescope, La Palma, yielding 
    $z = 0.685$ based on a single emission line adopted to be 
    [\ion{O}{2}] $\lambda$3727. This spectrum is mentioned at 
    http://www-astro.physics.ox.ac.uk/$\sim$sr/grimes.html.} 
  [\ion{Ne}{5}] (redshifted to $\sim$5560\AA) 
  was not detected in our Palomar spectrum, likely 
  owing to its proximity to the dichroic crossover.

  \subsubsection{Keck}
  \label{section_keck_spectrum}

  \begin{figure}[t]
    \includegraphics[width=\columnwidth,clip=true]{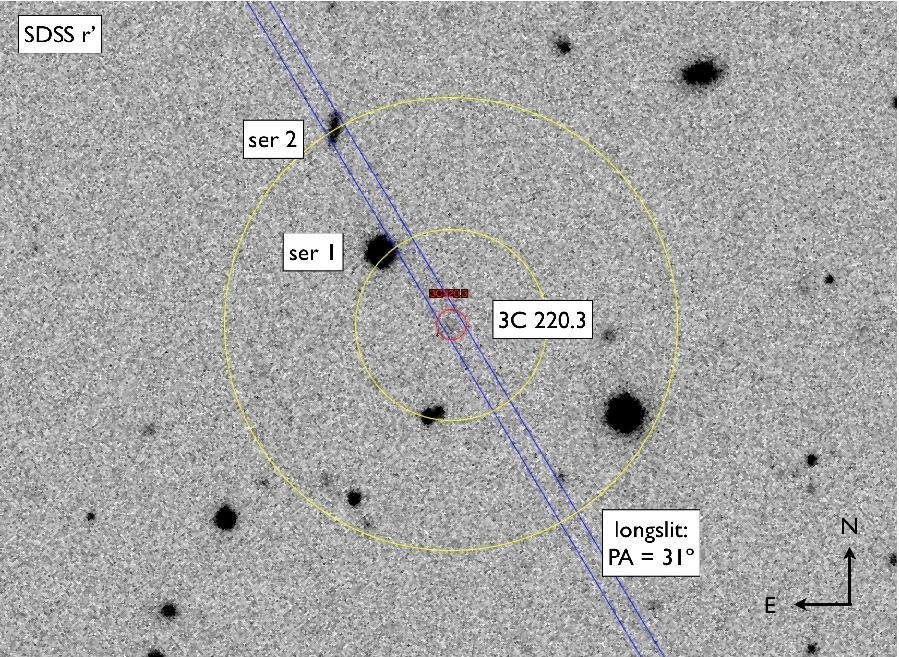}
    \caption{
      Keck LRIS slit position (blue lines) overlaid on an SDSS $r$-band
      negative image of the 3C\,220.3 system. 
      The small red circle marks the location of 3C\,220.3, 
      and the larger yellow circles (radii 19$\arcsec$ and 45$\arcsec$) 
      go through two serendipitously hit sources labelled ``ser 1'' and
      ``ser 2.'' 
    }
    \label{fig_slit_position}
  \end{figure}

  With the goal of measuring the arc redshift, 
  we performed additional spectroscopy of 3C\,220.3 
  with the Low Resolution Imaging Spectrometer (LRIS, 
  \citealt{Oke1995}) at the Cassegrain focus of the 
  Keck~I telescope. We obtained three 1200\,s 
  observations at a position angle of 31$^{\circ}$. 
  The observations used the 1\farcs0-wide longslit,  
  the 5600\,\AA\ dichroic, 
  the 400/3400 blue grism ($R \sim 900$), 
  and the 400/8500 red grating ($R \sim 1400$). 
  The position of the slit is illustrated in 
  Fig.~\ref{fig_slit_position}. 
  Conditions were photometric with $\sim$1\farcs5 seeing. 
  We processed the data using standard techniques 
  and flux calibrated the spectra using observations of 
  G191-B2B, a standard star from \citet{Massey1990}, 
  obtained during the same night.

  \begin{figure}[t]
    \includegraphics[height=\columnwidth,clip=true,angle=270]{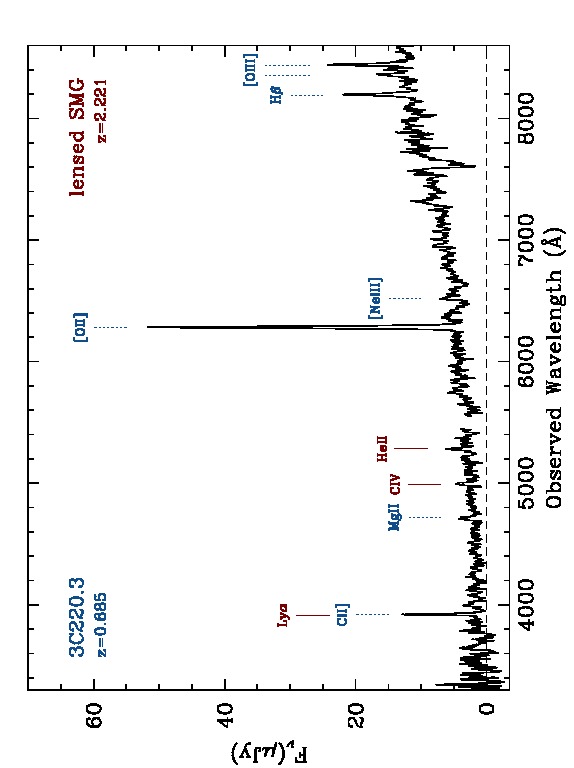}
    \caption{
      Keck LRIS spectrum of 3C220.3 system with prominent 
      lines labeled.  
      Blue dotted lines label emission lines associated with 
      the radio galaxy at $z = 0.685$, while red solid lines 
      label emission lines associated with the lensed SMG 
      at $z = 2.221$.  The feature at 3921 \AA\ is 
      associated with emission lines from both sources.  
      The gap at 5500 \AA\ corresponds to the dichroic 
      separating the blue and red arms of LRIS.  
      Absorption at 7600 \AA\ is due to the uncorrected 
      atmospheric A-band.  }
    \label{fig_keck_spectrum_all}
  \end{figure}

  \begin{figure*}
    \includegraphics[width=16cm,clip=true]{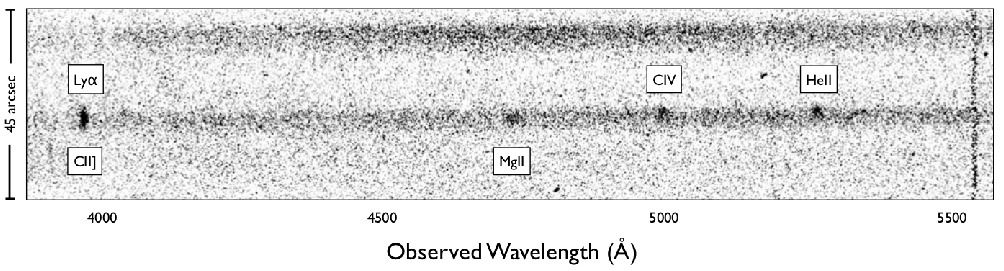}
    \caption{
      Negative image of the blue side of the Keck LRIS spectrum of 3C\,220.3. 
      The horizontal and vertical axes represent the spectral 
      and spatial dimensions, respectively. 
      The horizontal dark stripe in the middle of the frame 
      corresponds to the 
      3C\,220.3 system, and the upper dark stripe 
      originates from the faint extended source 
      labelled ``ser 1'' in Fig.~\ref{fig_slit_position}. 
      The 3C\,220.3 stripe has a width of $\sim$4\arcsec, 
      the lower part originating from source B, 
      the central part from source A, 
      and the upper part from the arc. 
      The dark knots in the stripe are due to emission lines 
      as labeled. \ion{Mg}{2} $\lambda$2798 from the radio galaxy 
      (source A) is centrally located in the continuum, 
      while the \ion{C}{4} and \ion{He}{2} lines are offset upwards, 
      indicating an origin in the arc. 
      The line on the left side (at 3920\,\AA) seems more central, 
      suggesting it is primarily \ion{C}{2} from the radio galaxy 
      rather than Ly$\alpha$ from the lensed SMG. 
    }
    \label{fig_keck_spectrum_blue}
  \end{figure*}

  The Keck spectrum confirms the redshift of the radio 
  galaxy, detecting the emission lines previously 
  seen in the Palomar spectrum as well as weak, narrow 
  [\ion{Mg}{2}] $\lambda$2800 and H$\beta$ emission 
  (Fig.~\ref{fig_keck_spectrum_all}). 
  More excitingly, the Keck spectrum reveals two additional 
  narrow emission lines with a slight spatial offset from 
  the primary continuum. 
  These lines are associated with the arc 
  (Fig.~\ref{fig_keck_spectrum_blue}), and
  we identify the lines as \ion{C}{4} $\lambda$1549 and 
  \ion{He}{2} $\lambda$1640 at redshift $z = 2.221$. 
  These lines are typical of radio galaxies and 
  type-2 AGN, and the high \ion{C}{4} luminosity
  (${\sim}2 \times 10^{42}$\,erg\,s$^{-1}$) argues for a 
  luminous AGN, i.e., a QSO (e.g., \citealt{Stern2002}). 

  In an odd coincidence, Ly$\alpha$ for the lensed 
  background source is shifted to observed 3916\,\AA, 
  which coincides with \ion{C}{2}] $\lambda$2326 from 
    the radio galaxy.
    The spatial location of this emission 
    in the two-dimensional spectrum 
    (Fig.~\ref{fig_keck_spectrum_blue})
    suggests that the emission at this wavelength 
    predominantly coincides with the radio galaxy; 
    Ly$\alpha$ emission from the lensed, background 
    SMG appears narrow and weak 
    (e.g., relative to the \ion{C}{4} line for an average 
    type-2 AGN), likely due to absorption and/or 
    resonance scattering by the dust- and gas-rich SMG.

    Examining the Keck spectra to uncover the redshift 
    of source B provides a less certain result. 
    The higher spectral dispersion red-arm data 
    were binned spatially  to 0\farcs27 per pixel 
    compared to 0\farcs135
    per pixel for the blue arm. 
    Based on both the [\ion{O}{2}] $\lambda$3727 and the 
    H$\beta$ emission lines, it appears that source 
    B is at nearly the same redshift as source A but blue-shifted by
    $\sim$5\,\AA\ (4~pixels). 

    \subsection{{\it Chandra\/} X-ray Observations}

\begin{figure}[t]
      \includegraphics[width=\columnwidth,clip=true]{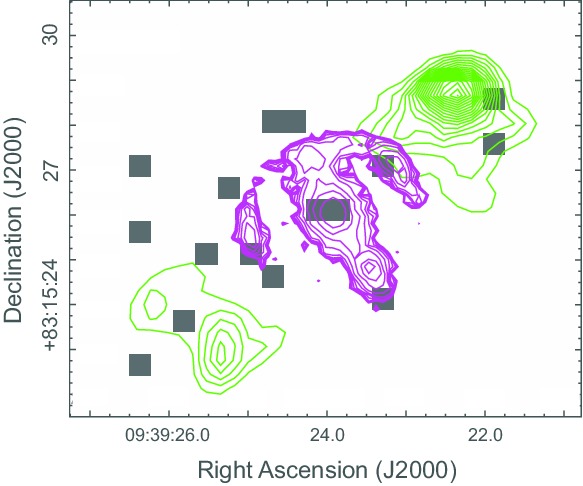}
      \caption{
        Chandra X-ray photon image (black squares) towards 3C220.3, with 
        Keck $K^{\prime}$-band contours (violet) and 
        6\,cm radio contours (green, from \citealt{Mullin2006}).
      }
      \label{fig_chandra}
    \end{figure}

    3C\,220.3 was observed with Chandra ACIS-S on 
    UT 2013 January 21 for 10\,ks. 
    The total, broad-band (0.3-–8\,keV) 
    X-ray flux density of the system is 
    ${\sim}1.14 \times 10^{-14}$\,erg\,cm$^{-2}$\,s$^{-1}$, 
    which yields a luminosity of 
    $L_{X} = 2.4 \times 10^{43}$\,erg~s$^{-1}$ if all photons 
    originate at $z = 0.685$. 
    The detection includes 16 photons (Fig.~\ref{fig_chandra})
    with 1.8 expected from a smooth background.  Photons are 
    distributed across the system. It is not generally
    possible to identify the X-ray photons with specific structure(s),
    although 2 photons are aligned with
    the radio core (implying $L_{X} \sim 3 \times 10^{42}$\,erg~s$^{-1}$). 
    The observational upper limit 
    of ${\sim}10^{43}$\,erg\,s$^{-1}$ for extended, hot X-ray 
    emitting gas is about a factor of ten 
    below typical values for X-ray clusters (\citealt{Pratt2009}).

    \subsection{Spectral energy distribution}

    \begin{figure}[t]
      \includegraphics[width=\columnwidth,clip=true]{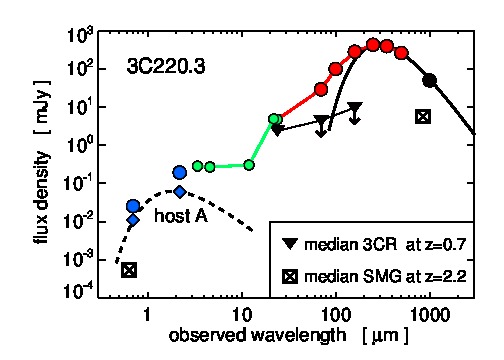}
      \caption{
        Spectral energy distribution of the 3C\,220.3 
        system. 
        Filled circles show flux densities from the entire 
        source: large red (\Her), small green 
        ({\it WISE\/} and \Sp), large black (SMA), 
        large blue (\HST\ and Keck). 
        A solid line shows a greybody fit through the SMA and 
        $\ge$\,250\,$\mu$m \Her\ data points 
        ($T = 36$\,K at $z = 2.221$, 
        dust emissivity index $\beta = 1.5$). 
        Blue diamonds show the \HST\ and Keck flux densities 
        of lens A (defined in Fig.~\ref{fig_high_res_images}) 
        connected by a blackbody (dashed line). 
        Triangles show the median 24--160\,$\mu$m 
        flux densities of twelve 3C radio galaxies at $z \sim 0.7$. 
        X-squares show the median $R$-band and 850\,$\mu$m 
        flux densities of 73 SMGs at $z \sim 2.2$ 
        (\citealt{Chapman2005}). 
      }
      \label{fig_sed}
    \end{figure}

    Figure~\ref{fig_sed} and Table~\ref{table_photometry} give  the spectral 
    energy distribution (SED) of the combined objects.
    The \Her\ beam  includes all flux from the 
    SMG knots and arc and from the two lenses A and B, but the
    contributions of A and B are likely to be negligible.
    In the comparison sample of 12 \Her-observed 
    3C objects,
    only three  were detected with \Her. 
    Their SEDs peak at
    $\lambda_{obs} \sim 100\,\mu$m  with flux densities below 100\,mJy  and
    decline towards shorter and longer wavelengths.
    The remaining nine 3C radio galaxies are undetected 
    at 70--160\,$\mu$m with upper limits well below the 
    3C\,220.3 flux densities. 
    Lens galaxy B is even less likely than A to contribute  to the 
    FIR flux density because it is $\sim$3 times fainter 
    than A in the NIR and has a slightly bluer visible-NIR 
    color (Table~\ref{table_a_b}, 
    $f(2.124\,\micron) / f(702\,{\rm nm}) \sim 5$ and 4 for A and B, 
    respectively). 
    The small 70--160\,$\mu$m flux densities of the 
    12 comparison 3C objects and the non-detection of A and B 
    at 1\,mm imply that most of the $\lambda>$200\,$\mu$m  flux
    comes from the SMG at $z = 2.221$. 

    The visible-to-850 $\mu$m flux density ratio of the 3C\,220.3 SMG is 
    consistent with the median of 73 SMGs at $z \sim 2.2$ 
    (\citealt{Chapman2005}) (Fig.~\ref{fig_sed}). 
    The best-fit dust temperature $T \sim 36$\,K
    of the $250-1000\mu$m dust emission
    (for a dust emissivity index $\beta=1.5$ and $z=2.221$) 
    is typical for unlensed SMGs 
    (\citealt{Magnelli2012}).\footnote{SED 
      fitting with several dust components covering a range 
      of temperatures does not result in a unique stable 
      solution, but the temperature 
      $T$ of the FIR-mm decline is a quite stable, widely used 
      characteristic parameter.} 
    If the dust were associated with the radio source
    at $z = 0.685$, 
    its temperature would be $<$20\,K and its dust mass, 
    $M_{d}>10^{11}M_{\odot}$, would be similar to the stellar mass,
    cooler and more massive than typically observed in AGN 
    ($>$30\,K, $M_{d} < 0.01 M_{stars}$). 
    Accounting for a total magnification factor $\mu \sim 10$ 
    derived from lens modeling 
    (Sect.~\ref{section_lens_modeling}), the intrinsic 
    (de-magnified) FIR luminosity of the  SMG 
    is $L({\rm FIR)} \sim 1 \times 10^{13}$\,\Lsol.

    \section{Lens modeling}
    \label{section_lens_modeling}

    To derive the lens model of the
    3C\,220.3 system, we used the Keck AO image 
    because it has higher signal-to-noise ratio and 
    shows more details than the {\HST} and SMA images
    (Fig.~\ref{fig_high_res_images}). 
    The lens parameters derived from 
    the Keck image were then used to model 
    the {\HST} and SMA images. 

    We first tried a model with a single lens centered on galaxy A, 
    assuming object B is part of the lensed source.
    This failed because it leaves
    bright residuals at the B position. 
    In other words, B is too bright relative to the other lensed 
    features in the Keck image to be fully 
    explained as part of the lensed image.
    Moreover, B is not seen in the SMA image. 
    The (hypothetical) case that B is a low-luminosity 
    foreground galaxy with mass too low to influence the lensing 
    leaves bright residuals at the B position.
    The remaining possibility is that B is 
    part of the lens. 
    Therefore, our preferred model has two lenses 
    centered on galaxies A and B. 

    \begin{figure}[t]
      \includegraphics[width=\columnwidth,clip=true]{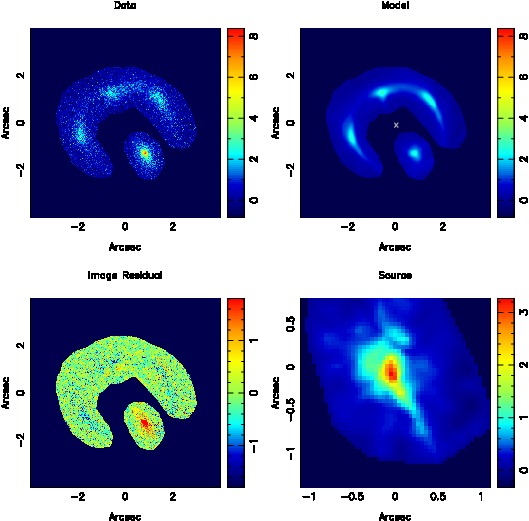}
      \caption{
        Results from modeling the Keck AO image with 
        a single lens centered on galaxy A, 
        located at (0,0). 
        The regions outside the Einstein half ring and source B are masked 
        and not used for the model fit. 
        While the total map has 200$\times$200 = 40000 pixels, 
        the area used for modeling has 10912 pixels.
        North is up and east is to the left. 
        {\it Top left:} Data after subtraction of A.
        {\it Top right:} Modeled image. 
        The cross marks the position of A.
        {\it Bottom left:} Residual image, data minus model image.
        {\it Bottom right:} Reconstructed image of the 
        source in the source plane         
        (50 $\times$ 50 = 2500 source pixels).
      }
      \label{fig_single_lens}
    \end{figure}

    \begin{figure}[t]
      \includegraphics[width=\columnwidth,clip=true]{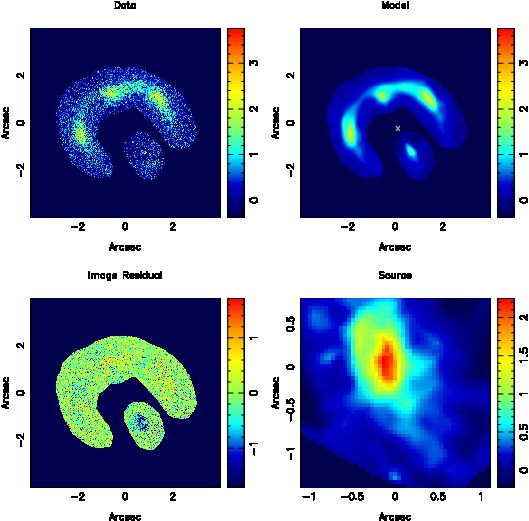}
      \caption{
        Results from modeling the Keck AO image with 
        a single lens centered on galaxy A, 
        as for Fig.~\ref{fig_single_lens}, 
        but using data after subtraction of both galaxies A and B.
      }
      \label{fig_ab_subtracted_single_lens}
    \end{figure}

    \subsection{Keck AO image}

    Lens models were derived from 
    a fully Bayesian, grid-based modeling technique 
    (\citealt{Koopmans2005}, \citealt{Vegetti2009}) 
    that simultaneously solves for 
    the parameters $\mathbf{\eta}$ of the lensing galaxies and 
    the surface brightness distribution of the background source 
    $\mathbf{s}$.  The Bayesian prior was expressed as the source 
    regularization form $\mathbf{R}$ with regularization level 
    $\lambda$ (\citealt{Vegetti2009}). 
    Because this prior is flat in 
    $\log(\lambda)$ and $\mathbf{\eta}$, the maximum of 
    $P(\mathbf{d}| \lambda, \mathbf{\eta}, \mathbf{M}, \mathbf{R})$, 
    that is the probability of the data $\mathbf{d}$ given 
    a lens mass model $\mathbf{M}$, determines the most 
    probable {\it a posteriori} fit 
    (see equation~35 of \citealt{Vegetti2009}).  
    The Bayesian approach takes into account the smoothness 
    of the source via the regularization 
    $\mathbf{R}$ at the level $\lambda$ (\citealt{Suyu2006}).
    For model fitting we have used the uncertainty map computed as 
    the
    quadrature sum of the image read out noise and Poisson noise 
    using the number of detected photons. 
    Notably the fitting results are very similar 
    (within a few percent) 
    to those obtained when 
    using a uniform uncertainty determined from clean background areas 
    of the image.
    
    To avoid that the model is affected by image pixels unrelated
    to the source, 
    all model fitting was restricted to 
    the "smallest-possible" mask as shown in Fig.~\ref{fig_single_lens}.
    The mask was set by smoothing the images, 
    then making a S/N cut. 
    This ensures that structure, that is present but below the noise
    level, is not cut off from the data.
    Thus, the mask is a compromise between 
    being large enough to allow identification of 
    possible extended residuals and being 
    small enough to cover only the arc and the area around 
    B, where the fourth counter-image is expected.

    We here consider three scenarios. 
    In case 1, object B is part of the lensed source;
    in case 2, B is a low-luminosity 
    foreground galaxy with mass  too small to influence the lensing;
    and in case 3, B is part of the lens. 
    The lens modeling began with case 1, using  
    a postage stamp image centered 
    on the lens system but with the 
    lensing galaxy (A) subtracted using analytic models. 
    For cases 2 and 3 both 
    galaxies (A and B) were subtracted. 
    The surface brightness of the galaxies
    A and B is best fitted by PSF-convolved S\'ersic (n$\sim$1)
    profiles (and not by de Vaucouleurs profiles). 
    Their parameters are listed
    in Table~\ref{table_a_b}.

    The lens mass distribution was modeled as an 
    ellipsoid with a free power-law 
    slope for the density profile.
    In general, such 
    approximations provide good fits to describe 
    the combined luminous and dark mass profiles 
    (e.g., \citealt{Treu2004}, \citealt{Auger2009}).  
    The profiles were centered on
    the light distribution of A 
    (or A and B), allowing for a small positional deviation. 
    The parameters of the best fit single and double lens models 
    are listed in Table~\ref{table_sie}.
    
    Case 1:
    Figure~\ref{fig_single_lens} shows that the positions 
    of the three knots connected by the arc can roughly be 
    fitted using the single lens A alone. 
    However, the image residual (data minus model) contains 
    not only a bright relict at the position of B but also
    negative features of the three 
    knots and the arc (seen as blue arcs in 
    the bottom left panel of Fig.~\ref{fig_single_lens}).
    In other words, B is too bright relative to the other lensed 
    features in the Keck image to be fully 
    explained as part of the lensed image.
    In addition, the light profile of B
    is well fit by a smooth S\'ersic profile that is unlike
    the complex structure of the three images
    along the arc, and finally the model requires the lensed source
    galaxy to be highly distorted.
    Overall case 1 is a poor fit to the data, and this is confirmed
    by the Bayesian evidence (Table~\ref{table_sie}).

    Case 2: 
    Figure~\ref{fig_ab_subtracted_single_lens} shows that
    the three knots connected by the arc 
    can be fitted, but there is a bright negative residual just 
    at the B position because this model requires a
    counter-image of the arc having about 2/3 the flux 
    of the three bright knots.
    Therefore, B should be seen in the SMA image, but it is not
    (Figs.~\ref{fig_high_res_images} and 
    \ref{fig_keck_lens_residuals}). 
    These discrepancies lead us to reject
    the single-lens model and consider B as part
    of the lens, consistent with its probable redshift 
    (Sect.~\ref{section_keck_spectrum}).
    
\begin{figure}[t]
      \includegraphics[width=\columnwidth,clip=true]{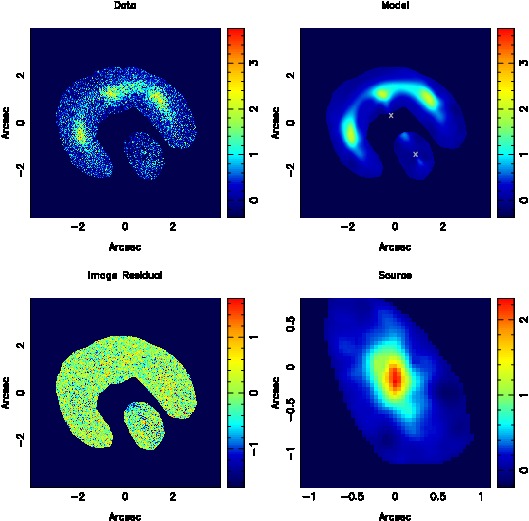}
      \caption{
        Results from modeling the Keck AO image with a double lens. 
        The regions outside the Einstein half ring and source B are masked 
        and not used for the model fit. 
        North is up and east is to the left. 
        {\it Top left:} Data after subtraction of A and B.
        {\it Top right:} Modeled image. 
        The crosses mark the positions of A and B. 
        There are two very faint objects, one between A and B and 
        one southwest of B; they may be the central image 
        and the counter-image of the arc.
        {\it Bottom left:} Residual image, data minus model image.
        {\it Bottom right:} Reconstructed image of the 
        source in the source plane.
      }
      \label{fig_double_lens}
    \end{figure}

\begin{figure}[t]
      \includegraphics[width=\columnwidth,clip=true]{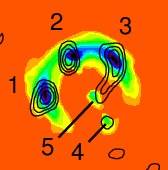}
      \caption{
        Modeled Keck AO image (color representing surface brightness) based on 
        the double lens. 
        Frame size is 7$\farcs$5.
        Superposed contours show the SMA map 
        with the three bright knots labeled. 
        The contours start at 2.5\,$\sigma$ and increase in steps of 
        1\,$\sigma$.
        Source image 4 coincides with a faint 1\,mm contour. 
        Source image 5 could be indicated by the faint tail of 
        knot 3. Galaxy B lies between knots 4 and 5.
      }
      \label{fig_keck_lens_residuals}
    \end{figure}

    Case 3: 
    Using two lenses significantly improves
    the modeling results.\footnote{The two-lens model includes the 
      possibility that B is a lens but not necessarily at the same redshift as
      A, because any two-lens-plane 
      system can be modeled to first order as a single lens plane with a 
      rescaled Einstein radius (\citealt{Keeton2001}).  The 
      Einstein radii are free parameters
      of the model, and therefore our results are independent of the true 
      redshift of B\null.  If its redshift is lower than assumed,
      its mass has to be higher, but all other model results are exactly
      the same.
    }
    Figure~\ref{fig_double_lens} illustrates the consistency of
    the observed and modeled images and the featureless 
    image residual all along the arc.
    The two-lens model predicts 
    two additional faint images: a fourth image on
    the opposite side of the Einstein half-ring 
    and a fifth (central) image. 
    Their positions agree with two faint blobs in the Keck data 
    after subtraction of a galaxy model for A and B and
    with faint features in the SMA map 
    (Fig.~\ref{fig_keck_lens_residuals}).
    However, both images are close to the noise level and 
    need to be confirmed with deeper optical/NIR data and 
    a higher spatial resolution 1\,mm map.
    The magnification factor of the two-lens model at 2.124\,$\mu$m is 
    $\mu \sim 7 \pm 2$.

    Visual inspection of the residual maps 
    (data $-$ model, Figs.~\ref{fig_single_lens}, 
    \ref{fig_ab_subtracted_single_lens} and~\ref{fig_double_lens})
    and the absence of B in the SMA image
    clearly favor the two-lens model.  
    The statistical quantification (Table~\ref{table_sie}) agrees.  
    The two-lens model has six extra free parameters 
    allowing us to fit the data better. 
    This is not an artifact of the higher number of free parameters, rather 
    the Bayesian evidence automatically encodes Occam's razor, 
    and models with more free parameters are automatically 
    penalized unless they are really required by the data.
    In essence, adding six extra parameters
    to the mass model 
    improves the Bayesian evidence by $\Delta\log Ev = 300$, 
    which corresponds to about 25-$\sigma$ difference,
    and produces a source galaxy that looks ``normal''
    (Fig.~\ref{fig_double_lens}).      
    If B were not important in the lensing, the model could have set
    its mass to zero, but this was not the 
    result.\footnote{
      We have used the Bayesian evidence to 
      determine the best-fit lens model, but for comparison 
      we also give here the
      more commonly used chi-square values. $\chi^{2}$ values were
      determined from the residual maps (data minus model) divided
      by the uncertainty map inside an area of 10912 pixels 
      (Fig.~\ref{fig_single_lens})
      However when source plane pixels
      are regularised, it is not straightforward to calculate
      the number of degrees of freedom $N({\rm dof})$.
      The calculation here excludes the 50 $\times$ 50 source pixels
      in the computation of $N({\rm dof)})$, which is therefore
      underestimated, thus overestimating the reduced $\chi^{2}_{r}$ .
      With 8 and 14 free parameters for the single- and 
      two-lens models $N({\rm dof}) = 10904$ and 10898 respectively.  
      The $\chi2$ values
      of cases 1, 2, 3 are 14393, 12976, and 12642, respectively.
      With these underestimated $N({\rm dof})$, 
      the $\chi^{2}_{r}$ values of cases 1, 2, 3 are 
      1.32, 1.19 and 1.16, respectively. 
      In the limit of a large number of degrees of freedom, 
      $N({\rm dof}) \gtrsim 10000$,
      $\chi^{2}_r$ has a standard deviation of 
      $(2/N({\rm dof}))^{0.5} \approx 0.014$, 
      hence the $\chi^{2}_r$ values for
      case 2 and case 3 
      differ at the 2-$\sigma$ level and are about 10$\sigma$ 
      away from unity. 
      The $\chi^{2}_r$ statistic suggests that either a 
      more complex lens model may be needed,
      or more likely that $N({\rm dof})$ has been underestimated,
      or that the noise map is underestimated by about 5-10\%, 
      which could quite easily be the case. 
    }
    
    \begin{figure}[t]
      \includegraphics[width=\columnwidth,clip=true]{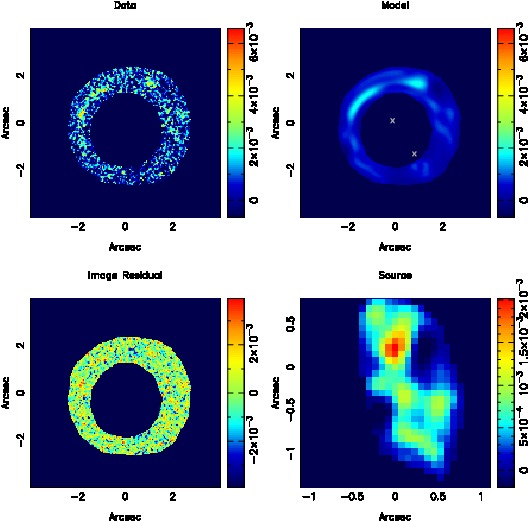}
      \caption{
        Results from modeling the {\it HST} 702\,nm image 
        with the double lens 
        model derived from the Keck AO image. 
        The area other than the Einstein ring is masked out. 
        North is up and east is to the left. 
        {\it Top left:} Data after subtraction of A and B.
        {\it Top right:} Modeled image. 
        The crosses mark the positions of A and B
        {\it Bottom left:} Residual image, data minus model image.
        {\it Bottom right:} Reconstructed image of the 
        source in the source plane.
      }
      \label{fig_double_lens_hst}
    \end{figure}
    
    Both the single and double lens models 
    require an external shear of $\sim$0.1  
    for lens A (Table~\ref{table_sie}), 
    suggesting the presence of additional lensing 
    components. 
    While the marginally visible knot 5 
    (Fig.~\ref{fig_keck_lens_residuals})  
    could be the fifth central image, 
    alternatively it could be a third 
    lensing galaxy C. 
    We checked for this posibility, 
    running a lens model with  three lenses
    centered on A, B and C.
    While C is still produced by the modeling as an extra image,
    a bit of mass is removed from B to C 
    (Einstein radius of C = 0$\farcs$030), 
    and the shear of lens A remains ($\gamma$=0.092).
    Also, the Bayesian evidence for the three lens model is 
    poorer than for the two-lens model, 
    and we do not use the three
    lens model further.
    Section~\ref{section_mass_to_light_ratio} discusses 
    the implications of the shear.
    
    \subsection{{\HST} image}

    Figure~\ref{fig_double_lens_hst} shows the results of applying the
    double lens model to the {\HST} image.  
    The image residuals are featureless, 
    supporting the adequacy of the
    model.

    \begin{figure}[t]
      \includegraphics[width=\columnwidth,clip=true]{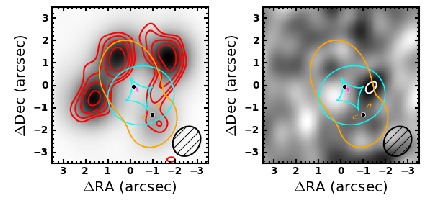}
      \caption{
        {\it Left:} SMA imaging (red contours, starting at 2$\sigma$ 
        and increasing by factors of 1.4) overlaid on the inverted, 
        deconvolved map of the best-fit model visibilities 
        (gray scale). 
        The critical curve and caustics are represented 
        by solid orange and cyan lines, respectively. 
        The position of the lenses and the source are shown 
        by plus signs and a magenta filled ellipse, respectively. 
        The FWHM of the synthesized beam is shown in the lower 
        right corner of each panel. 
        {\it Right:}
        Residual image (grayscale) obtained by inverting and 
        deconvolving the residual visibilities 
        (i.e., the difference between the model and data 
        visibilities). 
        Black and white contours indicate $\pm$2$\sigma$ level. 
        A 2$\sigma$ peak in the map of the 
        residual visibilities may be an indication of 
        additional minor structures in the lens plane or in 
        the source plane.
      }
      \label{fig_sma_lens}
    \end{figure}

    \subsection{SMA image}

    For interferometers such as the SMA, proper lens 
    modeling requires the use of visibilities in the 
    Fourier domain rather than surface brightness 
    maps. We followed the technique outlined by 
    \citet{Bussmann2012}. 
    Using the Gravlens 
    software\footnote{http://redfive.rutgers.edu/$\sim$keeton/gravlens/}, 
    we ray-traced the emission in the source plane 
    to the image plane for a given lensing mass 
    distribution and source morphology. 
    This image plane surface brightness map 
    was then used as input to MIRIAD's {\sc uvmodel}
    task, which computes the Fourier transform 
    of an image and matches the resulting 
    visibilities to the sampling of the 
    observed SMA visibility dataset. 
    We used the {\sc emcee} software package (for details, see
    \citealt{Foreman-Mackey2013}) to perform a Markov chain Monte Carlo
    sampling of the posterior probability density function (PDF) of 
    our model parameters.

    We modeled the SMA data
    using the lens parameters of the two--lens model derived
    from the Keck AO data (Table~\ref{table_sie})
    and a
    S\'ersic profile to represent the morphology of the SMG in the source
    plane. The SMA data (having lower spatial
    resolution than the KECK AO data) do not justify a more complex 
    model.  
    The best-fit source has a S\'ersic index 
    $n_{s}=1.38 \pm 0.46$, half-light radius 
    $r_{s}=1.26 \pm 0.38~kpc$,  
    axis ratio $b/a = 0.52 \pm 0.09$, and 
    $PA = 20^{\circ} \pm 17^{\circ}$.
    
    The best-fit result is shown in Figure~\ref{fig_sma_lens}.  
    Because the model fitting is performed in the Fourier domain and
    then transformed to the image domain, 
    small residuals in the image domain are not
    surprising.
    Modeling the SMA data with and without shear
    results in equally good fits.
    The magnification factor
    at 1~mm is $\mu = 10 \pm 2$. 

    \section{Reconstructed SMG morphologies}
    \label{section_smg_reconstruction}

\begin{figure}[t]
      \includegraphics[width=\columnwidth,clip=true]{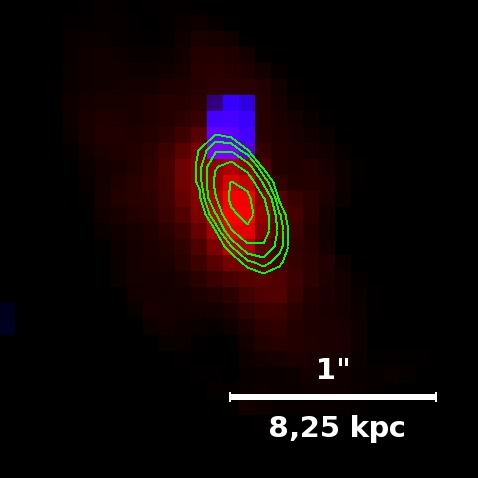}
      \caption{
        Source image of the SMG reconstructed 
        from the Keck, HST, and SMA images. 
        The stellar emission (Keck $\sim$680 nm restframe) 
        is shown in red, UV (\HST\ $\sim$218 nm restframe) 
        in blue, and the dust emission (SMA $\sim$310 $\mu$m restframe) 
        as green contours (logarithmic). 
        North is up and east is to the left.
      }
      \label{fig_reconstructed_smg}
    \end{figure}

    The difference in the observed morphologies of the SMG at 
    observed visible, NIR, and mm wavelengths 
    (Fig. \ref{fig_high_res_images}) 
    corresponds to morphological differences in the source at the 
    respective restframe wavelengths. 
    Fig.~\ref{fig_reconstructed_smg} shows the 
    multi-color image of the SMG in the source plane 
    as reconstructed from the modeling.
    Within astrometric errors, the peak of the dust 
    emission (SMA $\sim$310\,$\mu$m restframe) coincides with 
    the stellar light (Keck $\sim$680\,nm restframe). 
    However, compared to the main galaxy body of 
    stars and dust, the bulk of the UV ($\sim$218\,nm 
    restframe) emission is  shifted by 
    $\sim$4\,kpc (metric distance at the source) towards the north. 
    Marginal UV emission is also recognisable  
    $\sim$4\,kpc south of the galaxy body 
    (Fig.~\ref{fig_reconstructed_smg}, 
    and bottom right panel of Fig.~\ref{fig_double_lens_hst}). 
    These extended regions lie closer to the critical 
    curves of the lens model and may be somewhat more 
    magnified than the center of the stellar emission. 
    However, if a strong UV nucleus were present 
    within the compact dust emission, it should be visible
    in the observed \HST\ image as bright knots similar 
    to those in the SMA image. The \HST\ 
    image shows only the arc.

    The offset of the UV emission from the galaxy 
    body of the SMG suggests that the AGN responsible for the high
    excitation emission lines is 
    located in the galaxy center but hidden 
    from direct view at UV wavelengths. 
    The (apparent) bipolar morphology of the UV emission 
    is reminiscent of the ionisation cones seen in the 
    nearby starburst galaxy M82 
    (Fig. 3.108 of \citealt{Gil-de-Paz2007}), 
    in Seyfert galaxies (\citealt{Schmitt2003}), 
    and in high redshift quasars (e.g., \citealt{Cano-Diaz2012}). 
    One interpretation is
    that the SMG is breaking up its dust cocoon 
    so that UV radiation from the obscured 
    central region escapes through the opened 
    channels and is scattered towards us. 

    A luminous AGN  breaking up its dust cocoon is one
    interpretation of
    the so-called alignment effect 
    in powerful radio galaxies at high redshift 
    (\citealt{Chambers1987}, \citealt{McCarthy1987}).  In those cases, the
    breakup is apparently caused by the high-energy particles that create
    the radio lobes.
    In contrast to such radio galaxies, 
    the JVLA radio map of 3C\,220.3 shows no 
    indication for powerful radio activity 
    in the SMG, either at the known position 
    of lensed features (the arc and the three knots) 
    or from extended lobes, which would be expected 
    to be aligned with the UV emission 
    (approximately north--south). 
    While starburst winds may be able to blow holes 
    into the dust cocoon of the SMG and to expel 
    part of the gas (\citealt{Tenorio-Tagle1998}), 
    a wind from the hidden 
    QSO may be more likely to be forming the 
    open channels. Regardless of the mechanism, 
    the in-progress breakup
    of the dust cocoon makes the 3C\,220.3 SMG  
    an excellent lensed laboratory to explore, with 
    brightness magnification and at higher spatial 
    resolution than possible for unlensed systems, the 
    properties of a dust-enshrouded AGN at high redshift 
    and its feedback on the star forming activity in the host. 

    \section{The mass-to-light ratio of 3C\,220.3}
    \label{section_mass_to_light_ratio}

    The double lens model (Table~\ref{table_sie}) 
    gives 
    total -- baryonic and dark -- masses for
    the lensing galaxies A and B\null.
    The calculated masses are
    $M_{A} \sim 3.5 \times 10^{11}$\,\Msol\
    inside a cylinder of Einstein radius 
    $r_{E} = 1\farcs02$ ($= 7.2$\,kpc at $z = 0.685$), 
    and $M_{B} \sim 1.2 \times 10^{11}$\,\Msol\
    ($r_{E} = 0\farcs61 = 4.2$\,kpc).
    The statistical uncertainties (Table~\ref{table_sie}),  as determined 
    from the model residuals,
    do not account for possible systematic effects.

    One systematic effect for the mass estimates is that the best-fit model
    requires an external shear of $\sim$0.1  
    for lens A (Table~\ref{table_sie}), 
    suggesting the presence of additional lensing 
    components. This may be consistent with the presence of the 2$\sigma$ 
    residuals of modeling the SMA image 
    (Fig.~\ref{fig_sma_lens}).
    In principle, such a component could be a dark matter halo encompassing 
    A and B. On the other hand, our {\it Chandra} image shows no
    extended X-ray cluster gas (Fig.~\ref{fig_chandra}), 
    arguing against  explaining the shear 
    by a common dark matter halo. 
    More likely is that 3C\,220.3 is located in a small group with one or more
    faint galaxies having
    escaped detection on our images so far (possibly masked by the
    lensed structures). 
    A similar situation was 
    reported for the gravitational lens 
    CLASS B2045+265, 
    where the initially inferred high mass-to-light ratio 
    was reduced by a group of faint lensing satellites 
    (\citealt{Fassnacht1999}, \citealt{McKean2007}).
    Therefore we consider the masses $M_{A}$ and $M_{B}$ 
    as upper limits.\footnote{
      The single lens model of 3C\,220.3 yields 
      $M_{\rm tot}$ \,$\sim$ \,$9 \times 10^{11}$\,$M_{\odot}$, which is 
      a factor two higher than the sum of $M_{A}$ and $M_{B}$ 
      in the two--component lens model.
      If the shear of the two component lens model is
      due to further not yet identified lensing galaxies, 
      the mass sum may be reduced 
      further.} 

    To estimate the luminous mass of galaxies A and B 
    inside the Einstein radius, only two filters (Table~\ref{table_a_b})
    provide adequate data.
    (The spectra show no  stellar features.)
    Using the Starburst99 tool Version 6.0.4 
    (\citealt{Leitherer1999})\footnote{http://www.stsci.edu/science/starburst99/docs/default.htm} 
    and adopting a dominant old stellar population typical for radio galaxy hosts
    (age $\sim$3\,Gyr and Salpeter IMF), we obtain 
    stellar 
    masses 
    $M^{*}_{A} \sim 2.2 \times 10^{11}$\,\Msol, and 
    $M^{*}_{B} \sim 5.5 \times 10^{10}$\,\Msol.
    The uncertainty of these estimates is about a factor of two, 
    dominated by uncertainties in the photometry  of A and B 
    inside the Einstein radius (especially from the noisy 
    {\HST} image) and the unknown age
    of the stellar population.
    With these numbers, the ratio of dark-to-luminous mass 
    (projected within 
    the Einstein radius) lies in the range between 0.7 and 1.2, 
    i.e., dark matter fractions $f_{DM}$ of about 40\% and 55\% with
    large uncertainties ($\sim$30\%).

    \begin{figure}[t]
      \includegraphics[width=\columnwidth,clip=true]{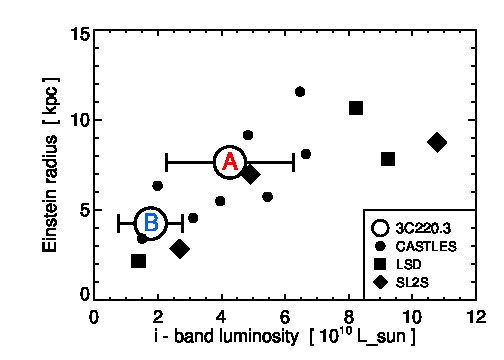}
      \caption{
        Einstein radius versus $i$--band luminosity 
        of
        3C\,220.3 A and B (large open circles) and the 
        CASTLES, LSD, and SL2S lensing galaxies
        at comparable redshift (filled symbols). 
      }
      \label{fig_reinstein_vs_lum_i}
    \end{figure}

    To check whether the radio galaxy 3C\,220.3 
    is associated with an usually high amount 
    of dark matter, suitable comparison data are required. 
    A proper comparison sample should contain field 
    galaxies of stellar mass and redshift similar 
    to 3C\,220.3. 
    The best existing comparison samples are 
    the CfA-Arizona Space Telescope LEns Survey of  
    gravitational lenses (CASTLES\footnote{Kochanek et al., 
      http://www.cfa.harvard.edu/castles} e.g., \citealt{Lehar2000}) 
    and a few $z \sim 0.6$ objects in the 
    the Lenses Structure and Dynamics survey 
    (LSD, \citealt{Treu2004}) and the 
    Strong Lenses in the Legacy Survey (SL2S, \citealt{Ruff2011}).
    Figure~\ref{fig_reinstein_vs_lum_i} shows the Einstein radius 
    (converted to kpc) versus the $i$-band 
    luminosity of the CASTLES, 
    LSD, and SL2S  
    lensing galaxies 
    in the redshift range $0.5<z<0.8$, 
    comparable to the redshift of 3C\,220.3, 
    and having lensed source redshift $z>1.3$ to ensure 
    that it is sufficiently far from the lens.
    The $i$-band luminosities of 3C\,220.3 A and B were 
    derived by interpolation of the {\HST} and Keck 
    photometry (Table~\ref{table_a_b}).
    Assuming the Einstein radius traces total mass, 
    the mass-to-light ratio of 3C\,220.3 A and B, 
    $M/L(i) \sim 8 \pm 4$, is comparable 
    to that of the CASTLES, LSD, and SL2S lenses, 
    perhaps at the high end of their distribution.
    LSD and SL2S list dark matter fractions 
    in the range 
    $0.25<f_{DM}<0.75$ 
    (projected within the Einstein radius), 
    encompassing the value 
    $f_{DM} \sim 0.4$ of 3C\,220.3.

    \section{Conclusions}
    \label{section_conclusions}

    The exceptional 3C\,220.3 gravitational lensing 
    system consists of a SMG (hosting a dust-enshrouded type-2 QSO)
    which is lensed by a powerful radio galaxy, i.e., 
    a radio-loud type-2 quasar. 
    The system permits  determination of the 
    total galaxy-scale mass for a powerful 
    double-lobed radio galaxy that is not 
    in a rich galaxy cluster. 
    Comparison with available galaxy-scale 
    lenses indicates an average to moderately 
    high dark matter fraction, but
    the comparison with radio-quiet galaxies needs to be confirmed 
    with future lens samples near the redshift 
    of 3C\,220.3. 
    Further observations of 3C\,220.3 with 
    increased $S/N$ will allow a more accurate 
    lens model, give a better stellar mass 
    measurement for the lenses, and allow a more 
    detailed reconstruction of the SMG morphology. 
    They should also clarify whether other galaxies important to the
    lensing, as for example in a small group, exist. 

    The SMG fits the paradigm of a composite 
    AGN/starburst galaxy in the early Universe. 
    The available data indicate the breakup 
    of the dust cocoon, which represents a 
    fundamental transition phase in a 
    violently star-forming galaxy. 
    Deeper X-ray observations and (sub)-mm 
    spectroscopy, both possible at high spatial 
    resolution, will allow exploration of 
    the break-up process and the 
    kinematics of any related outflows in a 
    high-redshift starburst-AGN.

    \acknowledgments
    We thank Shri R. Kulkarni, who enabled us to obtain the Keck LRIS spectrum, 
    and Dan Perley, who assisted with these observations. We also thank
    Dale Frail for granting Directors Discretionary time at the JVLA, 
    Adam Deller for helping with the JVLA data reduction, and
    Ray Blundell for granting Directors Discretionary time at the
    SMA\null. 
    We thank the referee for comments that improved the paper.
    \Her\ is an ESA space observatory with science instruments 
    provided by European-led Principal Investigator consortia 
    and with important participation from NASA\null. 
    The {\it Wide-field Infrared Survey Explorer\/} is a joint project of 
    the University of California, Los Angeles and the Jet 
    Propulsion Laboratory/California Institute of Technology. 
    This work is based in part on observations made with the 
    \SST, with the {\it Chandra X-ray Observatory\/}, 
    and on observations made with the 
    NASA/ESA \Hubble\ retrieved from the archive 
    at the Space Telescope Science Institute. 
    The National Radio Astronomy Observatory is a facility of 
    the National Science Foundation operated under cooperative 
    agreement by Associated Universities, Inc. 

    {\it Facilities:\null} \facility{Herschel}, \facility{HST}, \facility{Spitzer (IRAC, MIPS)}, \facility{EVLA}, \facility{SMA}, \facility{Keck:I (LRIS)}, \facility{Keck:II (NIRC2)}, \facility{Palomar (Hale)}, \facility{CXO}.
    \bibliography{refs}

    \clearpage 

    \begin{table*}
      \caption{
        Photometry of 3C\,220.3.
        \label{table_photometry}
      }
      \begin{tabular}{cr@{$\pm$}l}
        \tableline\tableline
        observed wavelength & \multicolumn{2}{c}{flux density} \\
        $\mu$m & \multicolumn{2}{c}{mJy} \\
        \tableline
        \00.702 & 0.0253&0.00 \\
        \02.124 & 0.193&0.03  \\ 
        3.6\0     & 0.275&0.028  \\
        4.5\0     & 0.283&0.028  \\
        70.\0\0\0 & 29.5&5  \\
        100.\0\0\0\0 & 102&7  \\
        160.\0\0\0\0 & 289&9  \\
        250.\0\0\0\0 & 440&15  \\
        350.\0\0\0\0 & 403&20 \\ 
        500.\0\0\0\0 & 268&30  \\
        1000.\0\0\0\0\0 & 51&12 \\
        \tableline
      \end{tabular}
      \tablecomments{Sum of all components A, B, knots and arc.}
    \end{table*}

    \begin{table*}[h!]
      \caption{
        Photometry of 3C\,220.3 A and B 
        and parameters of the surface profile modeling.
        \label{table_a_b}
      }
      \begin{tabular}{c|cc|cccc}
        \tableline\tableline
        Lens &  702 nm            &  2.124 $\mu$m  & S\'ersic Index $n$ & $r_{eff} [arcsec] $         & axis ratio $b/a$ & PA [deg]      \\
        \tableline
        A    &  14.8  $\pm$ 3.5   & 76.7 $\pm$ 15.0 & 0.9 $\pm$ 0.09 & 0.29 $\pm$ 0.04 & 0.7 $\pm$ 0.07 & 29 $\pm$ 12  \\
        B    &   4.45 $\pm$ 1.1   & 20.0 $\pm$ 3.0  & 1.0 $\pm$ 0.12 & 0.20 $\pm$ 0.04 & 0.6 $\pm$ 0.11 & 31 $\pm$ 25  \\ 
        \tableline 
      \end{tabular}
      \tablecomments{The flux densities in $\mu$Jy were determined
        using  SExtractor  with auto aperture option. 
        The resulting elliptical apertures have an average circular diameter 
        of about twice the Einstein radii of A and B.
        For surface profile modeling, the Keck AO data were fitted
        by PSF-convolved S\'ersic profiles.
      }
    \end{table*}

    \begin{table*}
      \caption{
        Model parameters from the Keck AO image.
        \label{table_sie}
      }
      \begin{tabular}{cccccccccc}
        \tableline\tableline
        (1)  & (2)         & (3)  & (4)  & (5)              & (6)       & (7)      & (8)       & (9)      & (10)  \\
        Lens & Einstein    & Lens position    & $b/a$& PA   & slope  & Shear    & Shear      & Lens mass   & ln Ev\\
        &radius [$\arcsec$]& $x,y$ [$\arcsec$]&      & deg  & s      &strength  & PA [deg]   & 10$^{11}$ \Msol    &    \\
        \tableline                         
        single lens, case 1 &          &  &  &  &  &  &       &     & -10341 \\
        A  & 1.67     &  0.043, -0.183 & 0.61 & 38.7 & 2.071 & 0.093 & 5.4  & 8.9  &  \\
        unc A  & 0.09 &  0.007,  0.010 & 0.10 &  8   & 0.140 & 0.031 & 20   & 1.1  &  \\
        \tableline\tableline                                                  
        single lens, case 2 &          &  &  &  &  &  &       &    & -9660  \\
        A      & 1.68 &  0.113, -0.324 & 0.27 & 30.8 & 2.310 & 0.146 & 13.3  & 9.0  &  \\
        unc A  & 0.08 &  0.013,  0.028 & 0.16 &  12  & 0.230 & 0.029 & 24    & 1.0  &  \\
        \tableline\tableline                                                  
        double lens, case 3  &          &  &  &  &  &  &       &    & -9409  \\
        A      & 1.02 & -0.174,  0.223 & 0.61 & 35.7 & 1.837  & 0.098 & 5.0 & 3.5  &   \\
        unc A  & 0.07 &  0.010,  0.010 & 0.13 & 10   & 0.090  & 0.030 & 23  & 0.5  &   \\
        B      & 0.61 &  0.861, -1.415 & 0.74 & 42.1 & 1.964  &       &     & 1.2   &  \\ 
        unc B  & 0.05 &  0.007,  0.008 & 0.12 & 15   & 0.150  &       &     & 0.2   &  \\ 
        \tableline
      \end{tabular}
      \tablecomments{
        The uncertainties (labeled ``unc'') are listed 
        in the line underneath the parameter values.
        The lens positions (in column 3) are relative to the axis 
        from the observer to 
        the source. 
        Columns 4 and 5 give the axis ratio ($b/a$ = minor/major axis) 
        and position angle of the lenses. 
        Radial mass surface density  $\rho \propto r^{-s}$ 
        (in column 6), a slope $s = 2.0$ corresponds to a 
        singular isothermal ellipsoid.
        Column 10 gives the natural logarithm of the Bayesian evidence 
        (ln Ev) of case 1,2,3, respectively.
      }
    \end{table*}

\end{document}